\documentclass[]{aa}
\usepackage{txfonts}
\usepackage{graphicx}
\usepackage{natbib}
\usepackage[utf8]{inputenc}

\begin{document}

\title{Young T-Dwarf Candidates in IC~348\thanks{Based on
    observations obtained with WIRCam, a joint project of CFHT,
    Taiwan, Korea, Canada, France, at the Canada-France-Hawaii
    Telescope (CFHT) which is operated by the National Research
    Council (NRC) of Canada, the Institute National des Sciences de
    l'Univers of the Centre National de la Recherche Scientifique of
    France, and the University of Hawaii.  Research supported by the
    Marie Curie Research Training Network "CONSTELLATION" under grant
    no. MRTN-CT-2006-035890}}

\author{A. S. M. Burgess\inst{1}
  \and E. Moraux\inst{1}
  \and J. Bouvier\inst{1}
  \and C. Marmo\inst{2}
  \and L. Albert\inst{3}
  \and H. Bouy\inst{4}
     }


\institute{Laboratoire d'Astrophysique,
Observatoire de Grenoble, CNRS, Universit\'e Joseph Fourier
B.P.  53, 38041 Grenoble Cedex 9, France
\and Institut d'Astrophysique de Paris, 98bis Bd Arago, 75014 Paris,
France
\and Canada-France-Hawaii Telescope Corporation, 65-1238 Mamalahoa
    Highway, Kamuela, HI 96743, USA
\and Instituto de Astrof\'{i}sica de Canarias, C/ V\'{i}a L\'{a}ctea s/n, E38205 - La Laguna (Tenerife), Spain}


\date{Received; Accepted }



\abstract
{The determination of the lower-end of the initial mass function (IMF)
  provides strong constraints on star formation theories.}
{We report here on a search for isolated planetary-mass objects in the
  3~Myr-old star-forming region IC~348. }
{Deep, narrowband $CH_4$off and $CH_4$on images were obtained with
  CFHT/WIRCam over 0.11 sq.deg. in the central part of IC~348 to
  identify young T-dwarfs from their 1.6~$\mu$m methane absorption
  bands.}
{We report three faint T-dwarf candidates with
  $CH_4$on$-CH_4$off colours \textgreater 0.4
  mag.  Extinction was estimated for each candidate and lies in the
  range A$_V\sim5-12$~mag. Comparisons with T-dwarf spectral
  models, and colour/colour and colour/magnitude diagrams, reject two of the three candidates because of their extreme $z'-J$ blueness.  The one remaining object is not thought to be a foreground field dwarf because of
  a number density argument and also its strong extinction A$_V\sim12$~mag, or thought to be
  a background field T-dwarf which would be expected to be much fainter.  Models and diagrams give this object a preliminary T6 spectral type.}
{With a few Jupiter masses, the young T-dwarf candidate
  reported here is potentially amongst the youngest, lowest mass objects
  detected in a star-forming region so far.  Its frequency is
  consistent with the extrapolation of current lognormal IMF estimates
  down to the planetary mass domain.}


\maketitle 


\section{Introduction}

To date more than 500 L-dwarfs and about 150
T-dwarfs\footnote{dwarfarchives.org - references within} have been
detected using a variety of optical and IR surveys. The vast majority
of detected brown dwarfs are old (3-6 Gyr) field dwarfs \citep{2009AJ....137....1F}, having cooled
over time.  The first detection of a T-dwarf in a star-forming region
was put forward by \cite{2002ApJ...578..536Z}, named \object{S Orionis
  70} and detected in \object{$\sigma$
  Orionis}. \cite{2004ApJ...604..827B} raises questions about the
authenticity of the object as a cluster member and suggested that
it was a field dwarf in the line of sight, which was backed up with a
statistical analysis. \cite{2004astro.ph.10678M} and
\cite{2008A&A...477..895Z} undertook further confirmation of the
object by follow-up membership and proper motion work. The population
of these young, low-mass objects is critical to furthering our
understanding of the very low-mass end of the initial mass function
(IMF). star-forming regions are well suited to search for the lowest
mass brown dwarfs as young objects are hottest, and so brightest
immediately after formation. Their temperature decreases over time, since
by their very nature they are not massive enough to begin fusion via
hydrogen or deuterium synthesis.

One ideal system in which to observe young brown dwarfs and constrain
the IMF is \object{IC~348}, a star-forming region towards the
direction of \object{Perseus}, centred at (J2000)
03$^{\mathrm{h}}$44$^{\mathrm{m}}$34$^{\mathrm{s}}$,
+32$^{\circ}$09$^{\prime}$8$^{\prime\prime}$, and embedded in the
foreground part of the \object{Per OB2 association}. The age of \object{IC~348}
has been determined to be $\sim 1-3$ Myr
\citep{2003AJ....125.2029M}. This star-forming region is relatively
nearby and whilst there is controversy surrounding its distance, IC
348 is located between 261$^{+27}_{-23}$ pc
\citep{1999A&AS..137..305S} and 316 pc \citep{1998ApJ...497..736H}
from the Sun whilst the \object{OB Per} association, of which
\object{IC~348} is a member, is taken to be between 315 pc
\citep{2003ApJ...593.1093L} and 340 pc \citep{1993BaltA...2..214C}. An
average value of 300 ($\pm$15 pc) has been used for this work
\citep{Herbst}. Extinction maps for \object{IC~348}, ranging from
$\sim$2{\textless}A$_V${\textless}20 mag depending on the cluster
region, were derived by \cite{1993BaltA...2..214C},
\cite{2003AJ....125.2029M}, and \cite{2006A&A...445..999C}.

\cite{2003AJ....125.2029M} derived the IMF of \object{IC~348} down to
35~M$_{Jup}$ and found it to be similar to the IMF of the
\object{Trapezium} cluster, having a mode between 0.1-0.2
M$_{\sun}$. Approximately 15-25\% of the population of the cluster
appear to be brown dwarfs and their spatial density is independent of
the distance from the cluster centre. The lowest mass objects were not
detected in their work because of the detection limits. Those objects
with a mass of a few M$_{Jup}$ are very faint and cool ($\sim$1200 K)
enough that methane can form in their atmospheres. The presence of
methane absorption bands in substellar objects defines the L/T dwarf
boundary \citep{2006ApJ...640.1063B}. This was first exploited by \cite{2003ApJ...597..555M} to
look for very faint objects in IC~348 to a depth of $H\sim$19.5 mag.
Narrow-band 1.65~$\mu$m methane imaging of \object{IC~348} was
conducted in order to filter out the lower mass T-dwarfs from the
L-dwarfs, by their $H-CH_4$ colour differences. They found 12-15
candidates which could be late M, L or T-dwarfs, with masses $\ge$ 5
M$_{Jup}$, although none have been confirmed spectroscopically because of
their faint nature. We report here on a new and deeper methane imaging
survey using 2 narrow-band filters, $CH_4$on and $CH_4$off, in order to better
distinguish between T-dwarfs and more massive young stellar objects
in \object{IC~348}.

\begin{table*}[t]
\begin{center}
\caption{Instruments used with field of view (FOV), filters,  and the central  
  coordinates of the pointings. The $J$, $H$ and $K_s$ final WIRCam images are
  made up of four overlapping fields (A, B, C and D). }
\label{table:surveys}
{\footnotesize      
\centering
\begin{tabular}{l l l l l}
\hline\hline
Telescope & FOV & Filter & Integration time (h) &Pointing coordinates (J2000)\\
\hline
CFHT MegaCam & 0.96{\degr}$\times$0.94{\degr} & $z^\prime$ & 2.5 & 03$^{\mathrm{h}}$44$^{\mathrm{m}}$36$^{\mathrm{s}}$.00 +32$^{\circ}$01$^{\arcmin}$50${\farcs}$0\\
CFHT WIRCam& 20 {\arcmin}$\times$20 {\arcmin} & $CH_4$off, $CH_4$on
&3.7, 1.4 & 03$^{\mathrm{h}}$44$^{\mathrm{m}}$14$^{\mathrm{s}}$.80 +32$^{\circ}$05$^{\arcmin}$06${\farcs}$0\\
CFHT WIRCam& 20 {\arcmin}$\times$20 {\arcmin} & $J$, $H$, $K_s$ &0.35, 0.16, 0.12 \\
&A&&& 03$^{\mathrm{h}}$44$^{\mathrm{m}}$14$^{\mathrm{s}}$.80 +32$^{\circ}$05$^{\arcmin}$06${\farcs}$0\\
&B&&& 03$^{\mathrm{h}}$44$^{\mathrm{m}}$14$^{\mathrm{s}}$.80 +32$^{\circ}$15$^{\arcmin}$06${\farcs}$0\\
&C&&& 03$^{\mathrm{h}}$45$^{\mathrm{m}}$02$^{\mathrm{s}}$.00 +32$^{\circ}$05$^{\arcmin}$06${\farcs}$0\\
&D&&& 03$^{\mathrm{h}}$45$^{\mathrm{m}}$02$^{\mathrm{s}}$.00 +32$^{\circ}$15$^{\arcmin}$06${\farcs}$0\\
\hline
\end{tabular}
}
\end{center}
\end{table*}

In Section 2 there appears a short description of the observations
and instrumentation used, and of the methane filters. We also describe
the photometric extraction method and the resulting catalogues. The
criteria for the selection of the candidates are presented in Section
3, and the promising candidates identified.  Section 4 discusses
the photometric properties of the candidates, their nature and
membership status, and Section 5 follows with conclusions.

\section{Observations and photometric extraction}

This work has made use of the 3.6m Canada France Hawaii
Telescope\footnote{www.cfht.hawaii.edu} (CFHT) on Mauna Kea,
specifically the infrared camera WIRCam \citep{2004SPIE.5492..978P} and the wide field imager MegaCam \citep{2003SPIE.4841...72B}.  Basic
information regarding these surveys is summarised in Table
\ref{table:surveys}, whilst the central wavelengths for the filters
used as well as the corresponding extinction coefficients can be found
in Table \ref{table:Av}. Central, {\itshape effective,} wavelengths and extinction coefficients are taken from the Spanish Virtual Observatory\footnote{http://svo.laeff.inta.es/theory/filters/index.php}.

\begin{table}[t]
\caption{Filters and central, effective, wavelengths used in this work.}
\label{table:Av}
\centering
\begin{tabular}{l l l l}
\hline\hline
Filter&$\lambda^{eff}_c$& $\Delta\lambda^{eff}$ & A$_{\lambda}$/A$_V$ \\
&$\mu$m&$\mu$m\\
\hline
  $z^\prime$   &   0.88 & 0.27  &   0.52 \\
  $J$    &  1.25 & 0.16  &  0.30 \\
  $CH_4$off  &   1.58 & 0.10  &   0.21 \\
  $H$    &  1.63 & 0.29  &  0.20 \\
  $CH_4$on   &   1.69 & 0.10  &   0.18 \\
  $K_s$    &  2.15 & 0.32  &  0.13 \\
\hline
\end{tabular}
\end{table}

\subsection{WIRCam}

WIRCam observations were obtained under programme 06BF23 in service mode
between September 2006 and January 2007.  A grid of four IR arrays of
2048$\times$2048 pixels make up the WIRCam detector yielding a field
of view of 20 {\arcmin}$\times$20 {\arcmin}. The pixel scale is 0.306
{\arcsec} or 0.15 {\arcsec} with microdithering (used for the $J$, $H$
and $K_s$ images). WIRCam was used to take narrowband $CH_4$off and
$CH_4$on along with $J$, $H$, and $K_s$ images of \object{IC~348}. The
$CH_4$on and $CH_4$off pointings are both centred at the same
location, which includes the cluster's center, while four overlapping
WIRCam fields were taken in $J$, $H$ and $K_s$ to provide a larger
areal coverage (see Table \ref{table:surveys}).  Total integration
times of 3.7h in $CH_4$off and 1.4h in $CH_4$on were obtained from 
multiple individual exposures of 30s, acquired using a 7-position
dithering pattern to fill in the 45$\arcsec$-wide gaps in between IR
arrays. The same observing strategy was used for JHK$_s$ images, with
individual integration times of 45, 10, and 15s, respectively,
yielding total integration times of 1260s ($J$), 560s ($H$) and 420s
($K_s$) for each pointing.

Individual images were detrended using the `i`iwi pipeline (Albert et
al., in prep.) at CFHT and sky subtraction, stacking, photometric and
astrometric calibrations, and quality control were performed at
Terapix\footnote{terapix.iap.fr} \citep{2007ASPC..376..285M}. The
seeing (PSF FWHM) measured on the final images was between
0.55{\arcsec} and 0.65{\arcsec}. The co-added JHK$_s$ images were
photometrically calibrated using the 2MASS catalogue over the same
area and renormalized to an arbitrary photometric zero-point of
30~mag. The methane images have no external photometric calibration
and the $CH_4$ magnitudes are given here on an arbitrary albeit
internally consistent scale, so that $CH_4$on-$CH_4$off$\simeq$0 for
unreddened field dwarfs (see below). The Terapix pipeline additionally
produced a catalogue of objects with aperture photometry for each
filter.

\subsection{MegaCam}

MegaCam observations were obtained under programme 06BF28 in queue
service mode on September 21-23, 2006.  MegaCam consists of a grid of 36
2k$\times$4k CCDs covering a 1{\degr}$\times$1{\degr} footprint. The
pixel size is 13.5 $\mu$m and the pixel scale is
0.185{\arcsec}. MegaCam was used to take 6 dithered $z^\prime$-band
images of 1500s each, yielding a total integration time of 9000~s.
After detrending the images at CFHT, sky-subtraction, stacking, and
photometric and astrometric calibrations were done at Terapix. The
images were photometrically calibrated using standard stars routinely
observed by the Queue Service Observing team at CFHT. The seeing
during the observations ranged from 0.65{\arcsec} to 0.80{\arcsec}.

\subsection{Photometric catalogues}

A combination of SExtractor \citep{1996A&AS..117..393B} and PSFex
(Bertin et al., in prep.) was used on each image to extract sources
and build a photometric catalogue for each image. On a first step,
SExtractor extracts well defined stellar-like objects, which are used
by PSFex to compute a PSF model that is allowed to vary with position
on the detector. Then, SExtractor uses this PSF model to more
accurately extract and measure the photometry of all the sources
detected on the image. Objects were detected at 3.5 $\sigma$ above the
background noise (see Section 2.4 for further details), most notably to avoid the high incidence of nebulous
emission and reflected nebulosity in the region. This threshold is
reasonably low because our interest lies in detecting the faintest
possible objects in the region, the methane T-dwarfs. Unfortunately,
this low threshold also increases the incidence of non-stellar-like
detections because of nebulosities, and many misdetections created via
crosstalk in the detector (though this was only particularly
problematic in the narrowband observations).

Further analysis was conducted using publicly-available 2MASS data for
those detected objects that were bright enough to have a 2MASS
counterpart. This was useful for two reasons: firstly that the
photometric accuracy and pipeline reduction of the WIRCam images could
be compared with those from 2MASS; and secondly that the difference in
the photometric systems could be estimated from the photometric
differences between the CFHT and 2MASS. The extracted $J$, $H$ and
$K_s$ catalogues were matched with the $J$, $H$ and $K_s$ 2MASS
catalogues resulting in 206 stars that had overlapping
magnitudes. Based on these, the dispersion between the extracted $J$,
$H$ and $K_s$ catalogues and the 2MASS $J$, $H$ and $K_s$ catalogues
was calculated. Good agreement was shown by mean magnitude differences
of 0.08 $\pm$ 0.02, 0.03 $\pm$ 0.05, and 0.04 $\pm$ 0.02 for $J$, $H$
and $K_s$ respectively. These values hold for the brightest stars in
our images, since fainter ones are undetected in the 2MASS survey. In
turn, this means that colour effects arising from CFHT filters are not
taken into account in this calibration. Indeed, while the photometric
zero-point used here is obtained from 2MASS, colour effects are not
corrected for and our WIRCam $J$, $H$, $K_s$, $CH_4$on and $CH_4$off photometry is given in the CFHT Vega system and MegaCam $z^\prime$ in the AB system.

\begin{center}
\begin{figure}[htbp]
\graphicspath{{images/}}
\begin{tabular}[c]{l}
\includegraphics[width=1.0\linewidth]{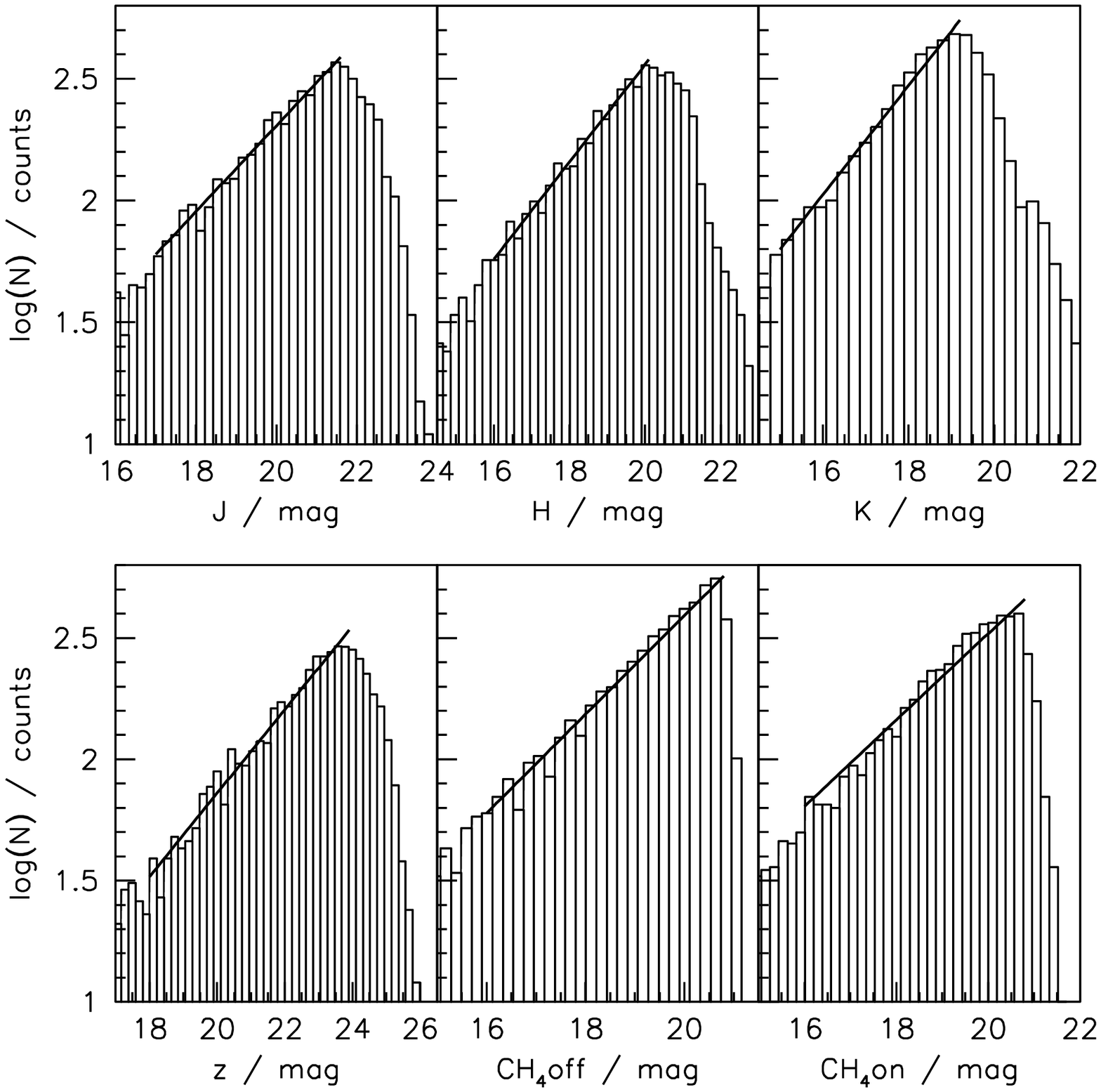}\\
\includegraphics[width=1.0\linewidth]{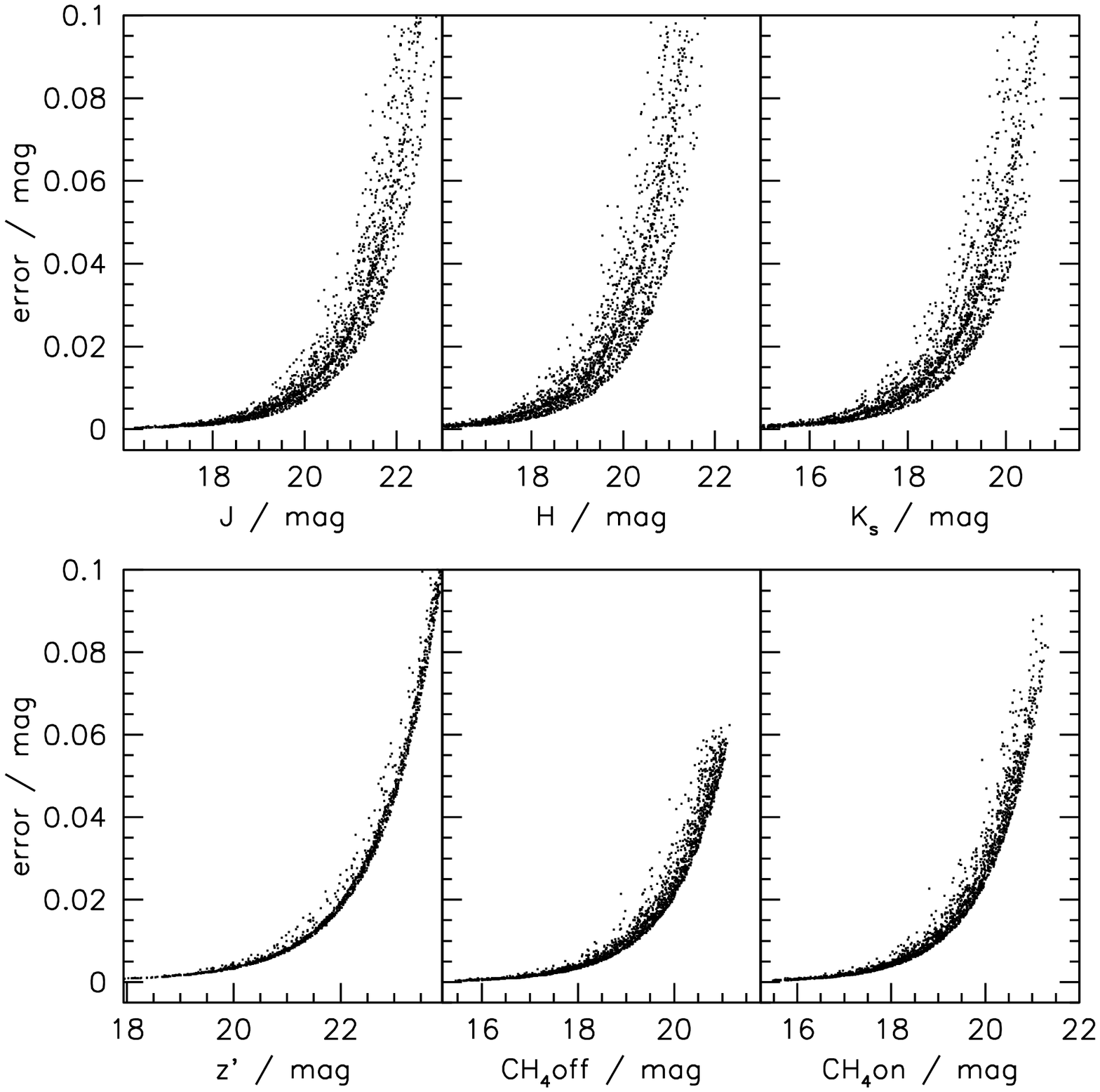}\\
\end{tabular}
\caption{{\itshape top}: log(N$_{Obj}$) vs magnitude (where N$_{Obj}$ is the number of detected objects in each band):  The completeness limits
  are found from these plots for the six imaged bands, $z$, $J$, $H$, $K$, $CH_4$on and $CH_4$off.  The line has been fitted to the histogram to find where the turning point occurs, thereby indicating our completeness limit. {\itshape bottom}: Photometric error, as measured by SExtractor, against magnitude.  Note that the $J, H, K_s$ images have multiple tracks as they are composed from 4 fields which were not acquired at the same time.}
\label{figtabupper}
\end{figure}
\end{center}

Each waveband had a catalogue created which was then collated and
cleaned of saturated objects, obvious artefacts, and a fraction of
extended sources identified from their large PSF FWHM.  An approximation of the completeness of the photometric catalogues was estimated from
$\log$(N$_{obj}$) vs. magnitude histograms, where N$_{obj}$ is the
number of stellar-like objects detected on the images. We thus derived
completeness limits from Figure~\ref{figtabupper} {\itshape(top)} of $\sim$23.5 ($z^\prime$), 21.5 ($J$), 20.0
($H$), 18.9 ($K_s$), 20.3 ($CH_4$on), and 20.7 ($CH_4$off).  SExtractor
photometric errors for each of the six bands are plotted in Figure~\ref{figtabupper} {\itshape(bottom)}, highlighting the increasing photometric error
for the faintening of the objects' brightness.  Uniquely, the $CH_4$off error terminates at 0.06 magnitudes as the $CH_4$off image was used for the detection of objects over all bands (see Section 2.4 for further details).

\subsection{Methane photometry}

\begin{figure}[t]
\setlength{\unitlength}{1cm}
\centering
\graphicspath{{images/}}
\includegraphics[width=1.0\linewidth]{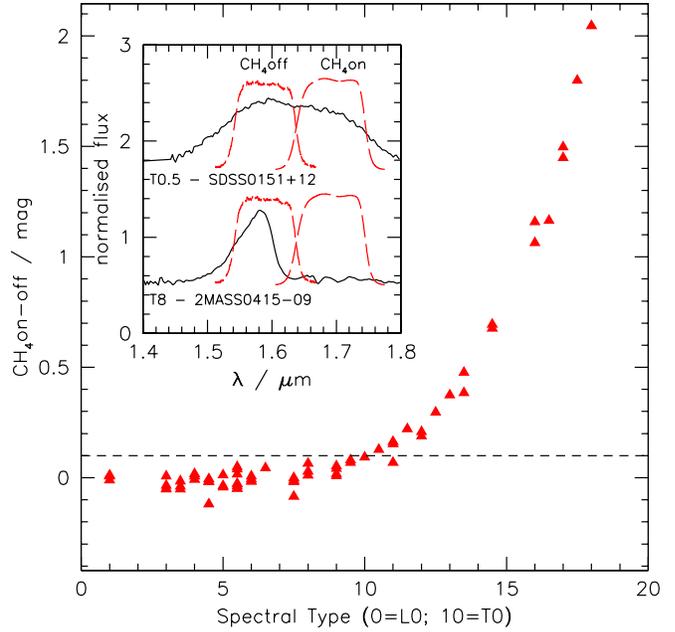}
\caption[$CH_4$on$-CH_4$off]{ $CH_4$on$-CH_4$off colour vs. spectral
  type. The triangles are computed methane colours for a spectral
  sequence of field L and T dwarfs (see text). The dotted line at
  $CH_4$on$-CH_4$off$=0.1$~mag highlights the limit where methane
  absorption becomes conspicuous in the photometric index, which
  occurs at a spectral type T0. Note how rapidly the methane index
  increases towards later spectral types.  {\it Inset}: Dwarf T0.5 and
  T8 spectra are shown in the region of the 1.6~$\mu$m methane
  absorption band.  The passband of the WIRCam $CH_4$off and $CH_4$on filters
  is superimposed (dashed). }
\label{figspt}
\end{figure}

The presence of methane absorption bands in the near-IR spectrum of
T-dwarfs can be used to identify T-dwarf candidates photometrically
(e.g. \citealt{2005AJ....130.2326T}).  Narrowband $CH_4$off
and $CH_4$on data was taken in order to classify these
objects. $CH_4$off measures the pseudo-continuum at 1.58~${\mu}$m
while $CH_4$on samples the methane absorption band at
1.69~${\mu}$m.  The passband of the two filters is overlain onto the
spectra of a T0.5 and a T8 dwarfs in Figure~\ref{figspt}.  Note the
greater methane absorption in the T8 dwarf spectrum compared to the
T0.5 spectrum in the region around 1.69 ${\mu}$m.

Spectra\footnote{www.jach.hawaii.edu/$\sim$skl/LTdata.html} of field
dwarfs from L1 to T8 were convolved \citep{008A&A...484..469D} with the WIRCam $CH_4$on and
$CH_4$off filters and the resulting methane colours plotted against
spectral type in Figure~\ref{figspt}.  The methane colours are seen to smoothly
increase towards later spectral types.  Whilst L-dwarfs have
$CH_4$on$-CH_4$off colours equal to zero, T-dwarfs have colours
above 0.1 mag which rapidly increase towards later T-types. To date, no other types of objects in current knowledge has this sequence of methane colours.  Thus, the
$CH_4$on$-CH_4$off colour provides a useful means to separate L and
T-dwarfs.  Additionally, as late-type T-dwarfs may remain undetected in the
$CH_4$on image, and due to strong $CH_4$ absorption, we first performed
object detection at $3.5\sigma$ on the $CH_4$off image, then performed PSF
photometry at the location of these objects on both the $CH_4$off and
$CH_4$on images.  This ensures that all objects that could possibly be
T-dwarfs are searched for in both filters. 

\begin{figure}[t]                                                                  
\setlength{\unitlength}{1cm}
\centering
\graphicspath{{images/}}
\includegraphics[width=0.9\linewidth]{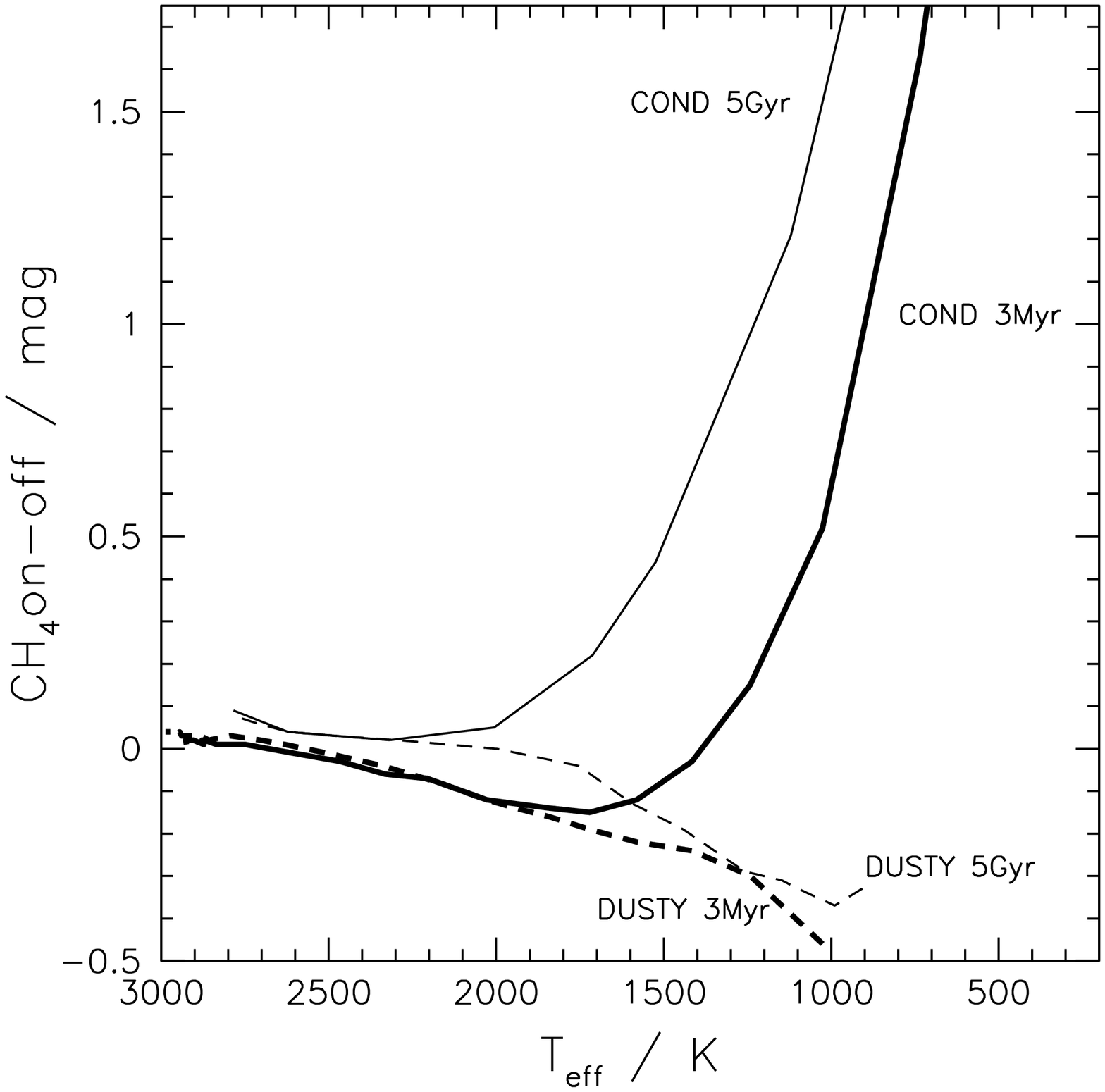}
\caption[T$_{eff}$ vs $CH_4$on$-CH_4$off]{ $CH_4$on$-CH_4$off colour
  vs. $T_{eff}$. 3 Myr COND {\it(thick solid)}, 5 Gyr COND {\it(thin
    solid)} 3 Myr DUSTY {\it(thick dashed)} and 5 Gyr DUSTY {\it(thin
    dashed)} models are shown. The COND models
  predict later spectral types for 3~Myr \object{IC~348} objects than for field
dwarfs for given methane colours.}
\label{figteff}
\end{figure}

Figure~\ref{figspt} provides an empirical calibration of the methane
colours against spectral type for field dwarfs.  For comparison,
Figure~\ref{figteff} shows the $CH_4$on$-CH_4$off colour as a function
of effective temperature ($T_{eff}$) as predicted by COND and DUSTY 3
Myr and 5 Gyr models
\citep{2000ApJ...542..464C,2001ApJ...556..357A,2003A&A...402..701B}.  While the DUSTY models are roughly similar at both ages, with
$CH_4$on$-CH_4$off $\sim$ 0 at $T_{eff}\geq 1500$~K, the COND models
predict bluer colours for younger objects at
$T_{eff}<1500$~K.  According to the models, \object{IC~348} T-dwarfs would then
have a smaller $T_{eff}$ and thus a later spectral type than field
T-dwarfs for the same $CH_4$on$-CH_4$off colour.

\section{Results}

Catalogues were obtained as described above for each of the
MegaCam/WIRCam images.  The COND and DUSTY 3 Myr and 5 Gyr models\footnote{http://phoenix.ens-lyon.fr/simulator/index.faces} are
also in the CFHT Vega system where the $z'$ band has been converted into AB magnitudes for our purposes. Similarly, the empirical field L-T dwarf sequence photometry has been adjusted to be consistant with the models and our data.  In the
colour/magnitude diagram presented the field dwarf sequence has been shifted to the distance of \object{IC~348}, taken to be at 300~pc.

\subsection{T-dwarf candidate selection}

\begin{figure}[t]
\setlength{\unitlength}{1cm}
\centering
\graphicspath{{images/}}
\includegraphics[width=1.0\linewidth]{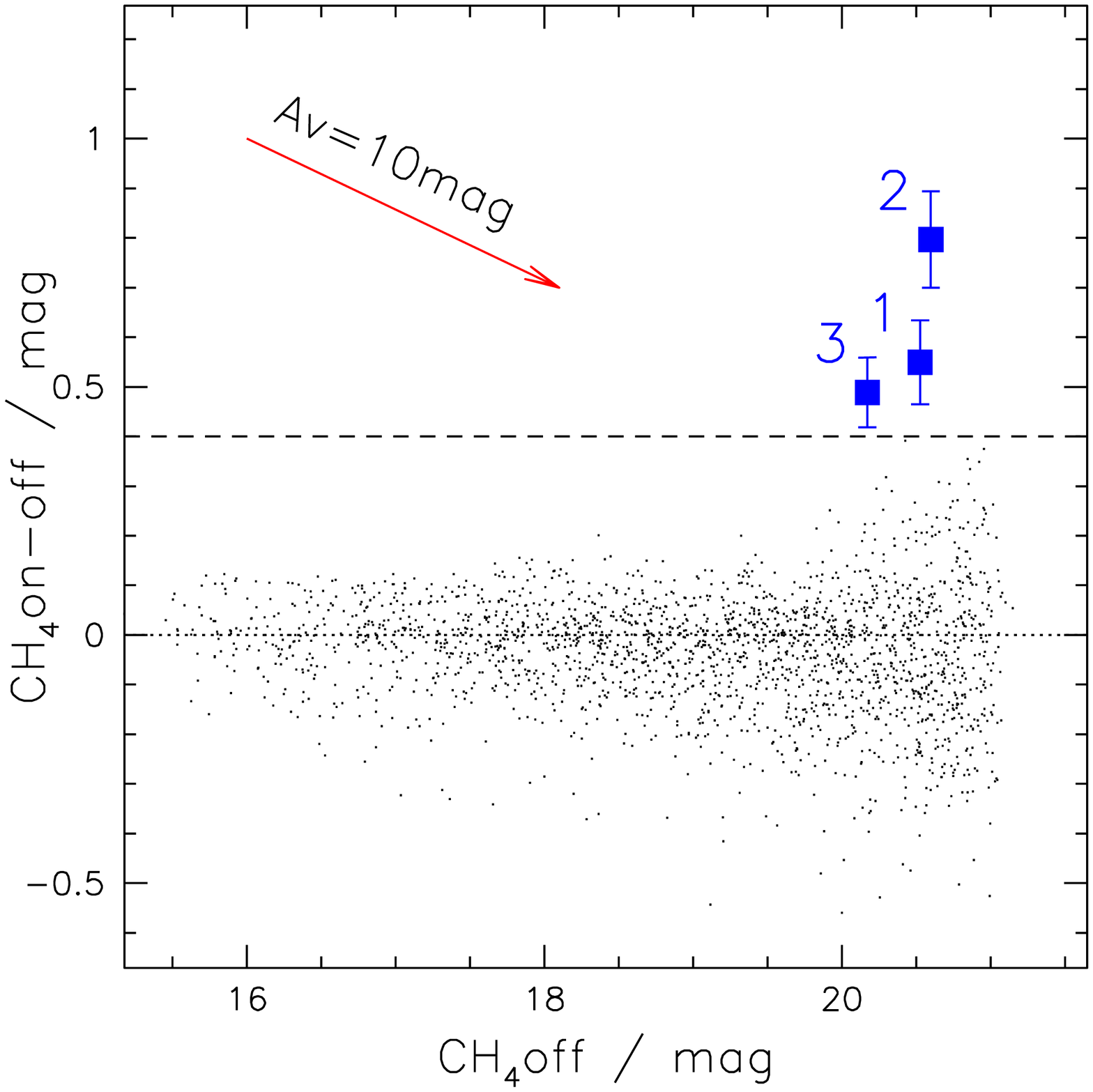}
\caption[$CH_4$on$-CH_4$offvs $CH_4$off]{$CH_4$on$-CH_4$off colour
  vs $CH_4$off.  The large (blue) squares show the three T-dwarf
  candidates with $CH_4$on$-CH_4$off $\ge 0.4$.  The extinction vector is
  shown for $A_V=10$~mag. The sources with strongly negative colours
  are deeply embedded objects (see text).}
\label{figOOO}
\end{figure}

Figure~\ref{figOOO} shows $CH_4$on$-CH_4$off against $CH_4$off for
stellar-like objects detected on WIRCam methane images.  An object was
considered as a T-dwarf candidate if $CH_4$on$-CH_4$off~$\ge$~0.4 mag,
which corresponds to $\geq$~3.5~${\sigma}$ above the L/T transition
($CH_4$on$-CH_4$off~$=0$) for the faintest detected objects.  Assuming
that young T-dwarfs follow the field dwarf methane relation, the
spectral type of those objects with $CH_4$on$-CH_4$off~$=$~0.4 mag
is close to T3 from Figure~\ref{figspt}.

\begin{center}
\begin{figure*}[t]
\graphicspath{{images/}}
\begin{tabular}[c]{lll}
\includegraphics[width=0.31\linewidth]{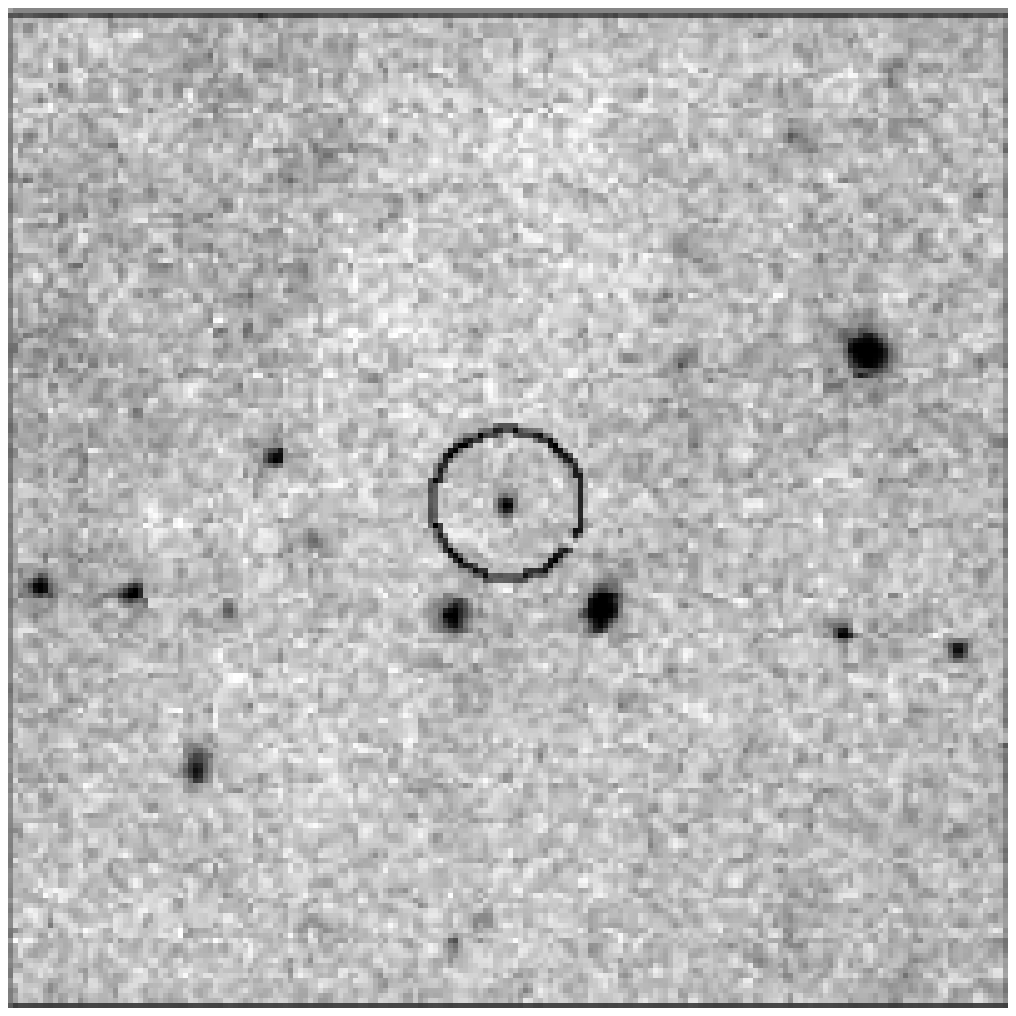} & \includegraphics[width=0.31\linewidth]{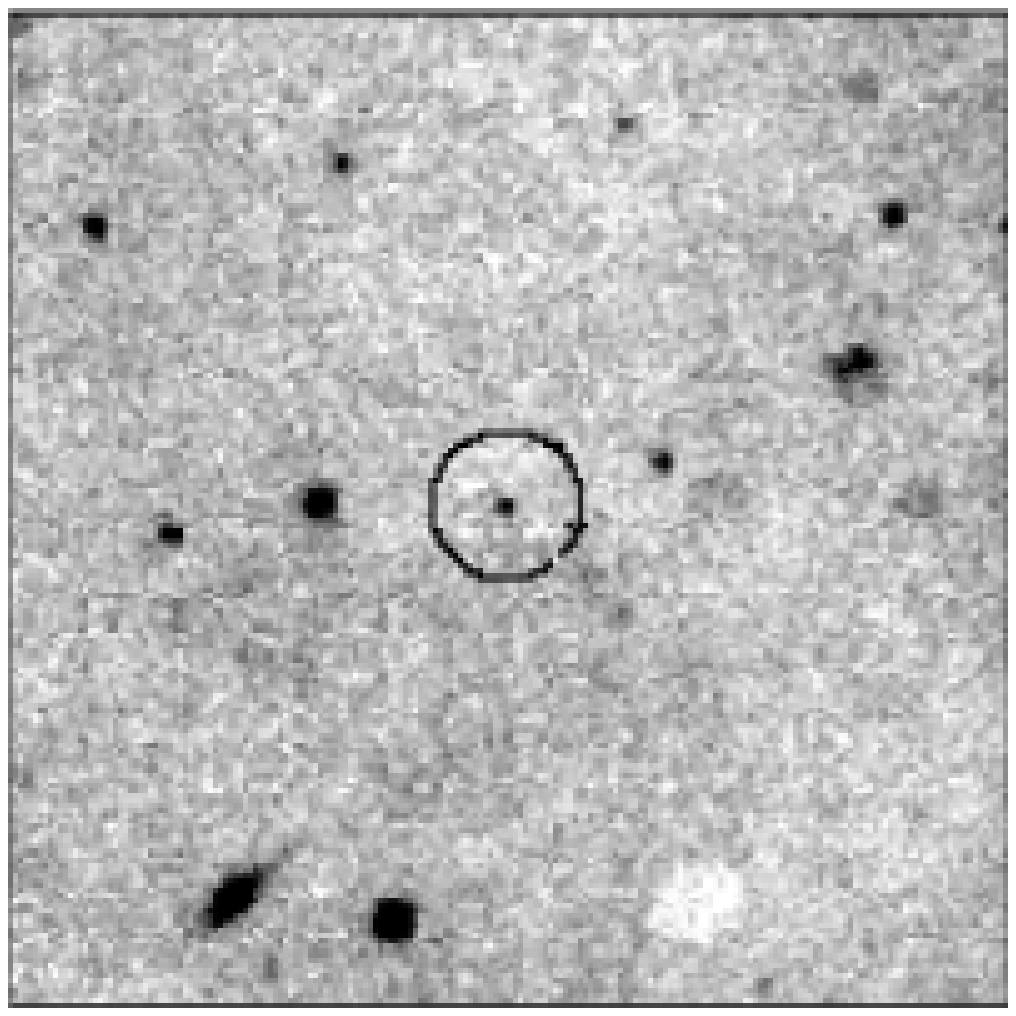} & \includegraphics[width=0.31\linewidth]{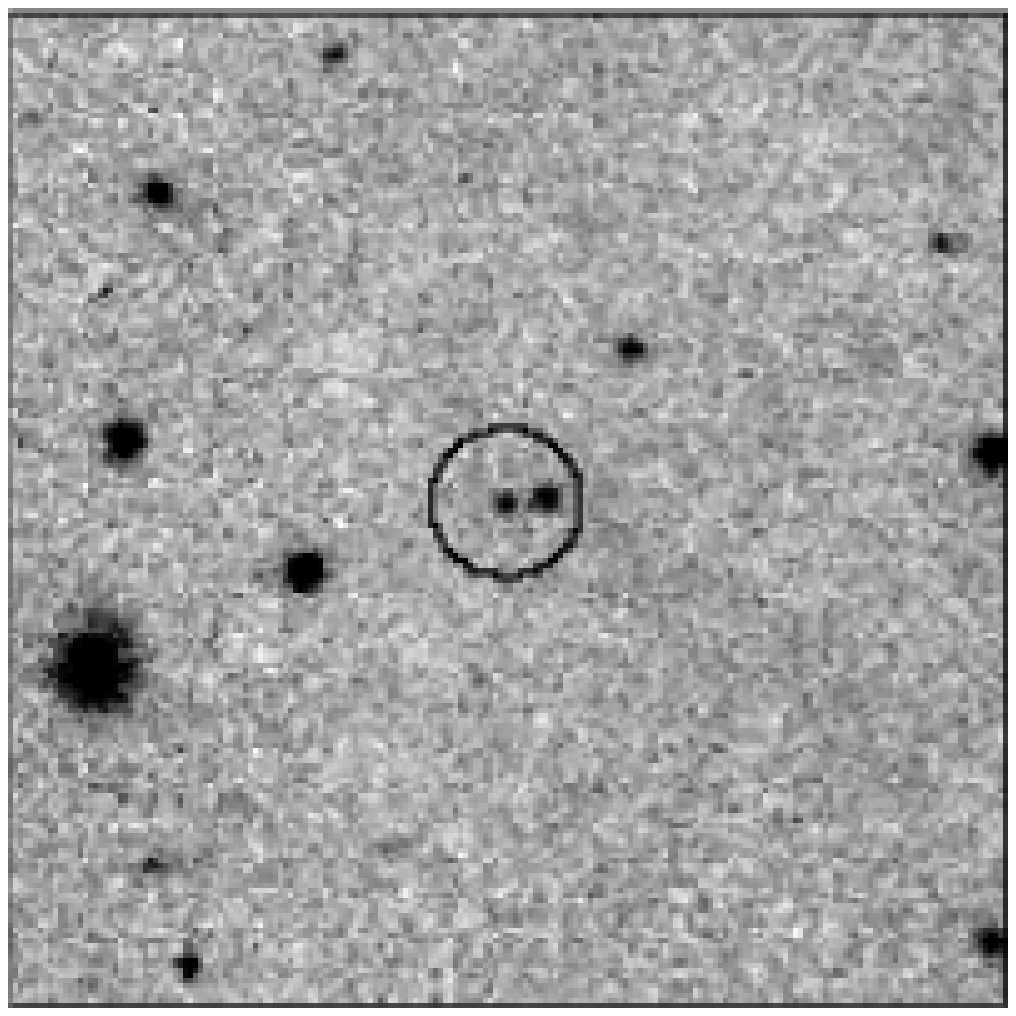}\\
\end{tabular}
\caption{Thumbsized $CH_4$off images of the three T-dwarf candidates,
  left to right, IC348\_CH4\_1, IC348\_CH4\_2 and IC348\_CH4\_3, respectively. Black circles
  highlight the location of the T-dwarf candidates. The images are
  48{\arcsec} to a side, where North is up and East is to the left.}
\label{figimages}
\end{figure*}
\end{center}

Objects with $CH_4$on$-CH_4$off~$\geq$~0.4~mag were initially all
classified as T-dwarf candidates.  The objects just below this limit
could still be T-dwarfs from their $CH_4$on$-CH_4$off colours as shown
in Figure~\ref{figspt}, but would be more difficult to extract because
of photometric errors.  With this criterion 136 sources were
selected.  This sample was then visually scrutinised on both the
$CH_4$off and $CH_4$on images before being reduced to a shortlist of
12 possible candidates.  Upon visual inspection, we found that the
rejected 124 `objects' were saturated stars, nebulous detections or
ghosts.  Of these 12 possible candidates, the shape of the PSF and the
contours were further examined using IRAF, where 9 were identified as
ghosts or detector cross-talk, so reducing this figure to 3 likely
candidates.  Thumbsized $CH_4$off images of the three candidates are
shown in Figure~\ref{figimages}.

The PSF photometry of the 3 candidates in z$^\prime$, $J$, $H$, $K_s$ and $CH_4$on/$CH_4$off
filters is listed in Table \ref{table:mags}. Candidate \object{IC348\_CH4\_2} was not
detected in the $z^\prime$-band. The $z^\prime$-band detection limit
was estimated by simulating a set of 1948 stars with magnitudes
between 20 and 26 using Skymaker \citep{2008SAIt..75..282B}. These synthetic stars
were then stacked onto the $CH_4$off footprint of the original
$z^\prime$ image using Swarp \citep{2007ASPC..376..285M}. At the 3$\sigma$ level, 1688 objects were detected where the
faintest had an input magnitude of 25.7, which we take as being the
detection limit in the $z^\prime$ filter.

\begin{table*}[t]
\caption{ PSF photometry and photometric errors of the three T-dwarf candidates, in magnitudes. }
\label{table:mags}  
\centering 
\begin{tabular}{l l l l l l l l l l l l l}
\hline\hline
 Object  &  $z^\prime$ & $\sigma_{z^\prime}$ & $J$ & $\sigma_{J}$ & $H$ & $\sigma_{H}$ & $K_s$ & $\sigma_{K_s}$ & $CH_4$off & $\sigma_{CH_4off}$ &  $CH_4$on &  $\sigma_{CH_4on}$\\
\hline
 \object{IC348\_CH4\_1}&  23.32      &   0.07  &  21.62 &  0.04  &  20.95 &  0.07  &  20.22 &  0.06  &  20.52 &  0.04  &  21.06 &  0.08 \\
 \object{IC348\_CH4\_2}&  $\ge$ 25.7 &   -     &  22.51 &  0.07  &  21.65 &  0.08  &  20.10 &  0.04  &  20.59 &  0.04  &  21.37 &  0.10 \\
 \object{IC348\_CH4\_3}&  23.85      &   0.10  &  22.02 &  0.07  &  21.02 &  0.07  &  19.94 &  0.05  &  20.17 &  0.03  &  20.66 &  0.06 \\
\hline                                                                                                                                                      
\end{tabular}
\end{table*}

Most of the objects detected on the WIRCam images lie along the line
$CH_4$on$-CH_4$off $=0$ in Figure~\ref{figOOO} as expected for field
dwarfs \citep{2005AJ....130.2326T}, the photometric error increasing
with magnitude. The 3 candidates are located in the faintest region of
the plot, whilst also having increasing methane colours. The
$CH_4$on$-CH_4$off rms photometric error at the magnitude of the
candidates is $\sigma$ $\sim$ 0.12 mag. The 3 candidates have
$CH_4$on$-CH_4$off colours of 0.54, 0.78, and 0.49~mag, respectively,
corresponding to a detection level of $>4\sigma$ (see
Table~\ref{table:top3}).

Note that a number of objects in Figure~\ref{figOOO} have strongly {\it
  negative} $CH_4$on$-CH_4$off colours, well beyond the rms
photometric error. Upon inspection of the images, all these objects
turn out to be young IC~348 members deeply embedded in bright compact
nebulosities. The large extinction they suffer results in strong
reddening, thus yielding blue (i.e. negative) $CH_4$on$-CH_4$off
colours in spite of the small wavelength difference between the 2
narrow-band methane filters. Reddening also accounts for the
asymmetric colour distribution of the faint background objects in this
plot.

Finally, a cross-comparison was made with the study of
\cite{2003ApJ...597..555M}. They classified 5 candidates as being
possibly M, L or T-dwarfs, all of which are detected in our images but
with $CH_4$on$-CH_4$off colours between -0.14 and 0.04~mag, placing
them out of our T-dwarf candidate criteria. 

\subsection{Reddening and spectral type estimates}

\begin{figure*}[t]
\setlength{\unitlength}{1cm}
\centering
\graphicspath{{images/}}
\begin{tabular}[c]{lll}
\includegraphics[width=0.31\linewidth]{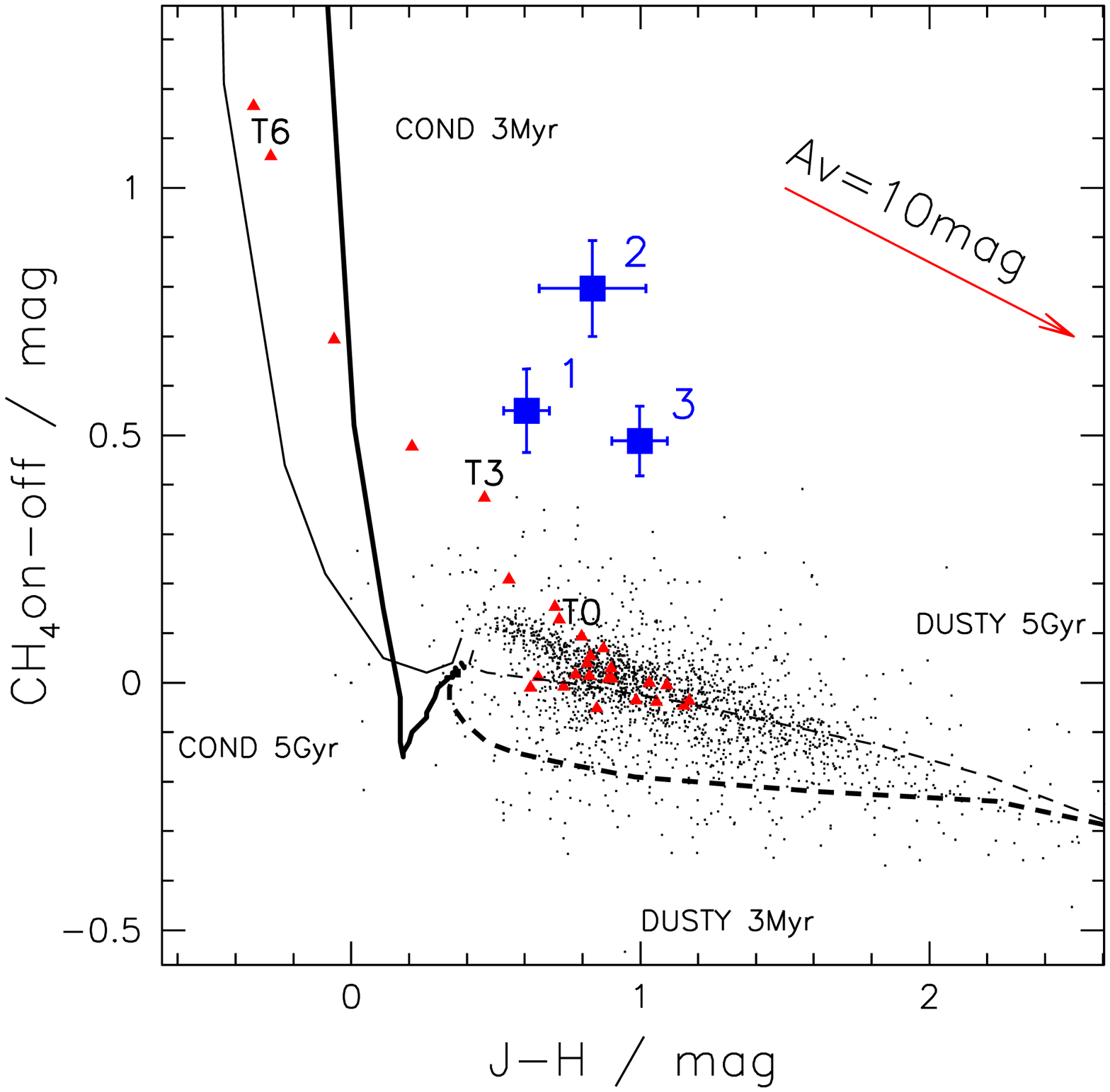} & \includegraphics[width=0.31\linewidth]{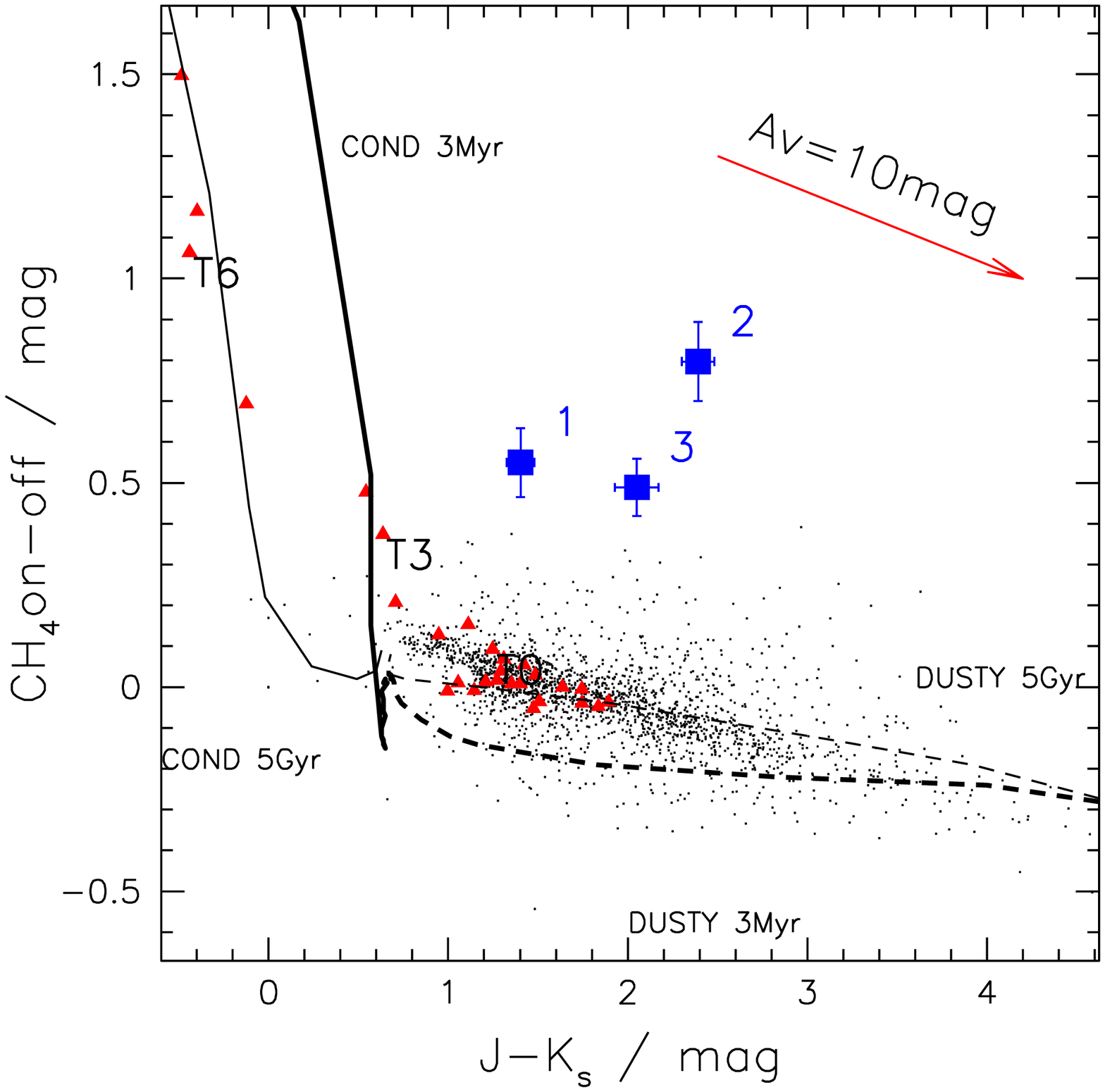} & \includegraphics[width=0.31\linewidth]{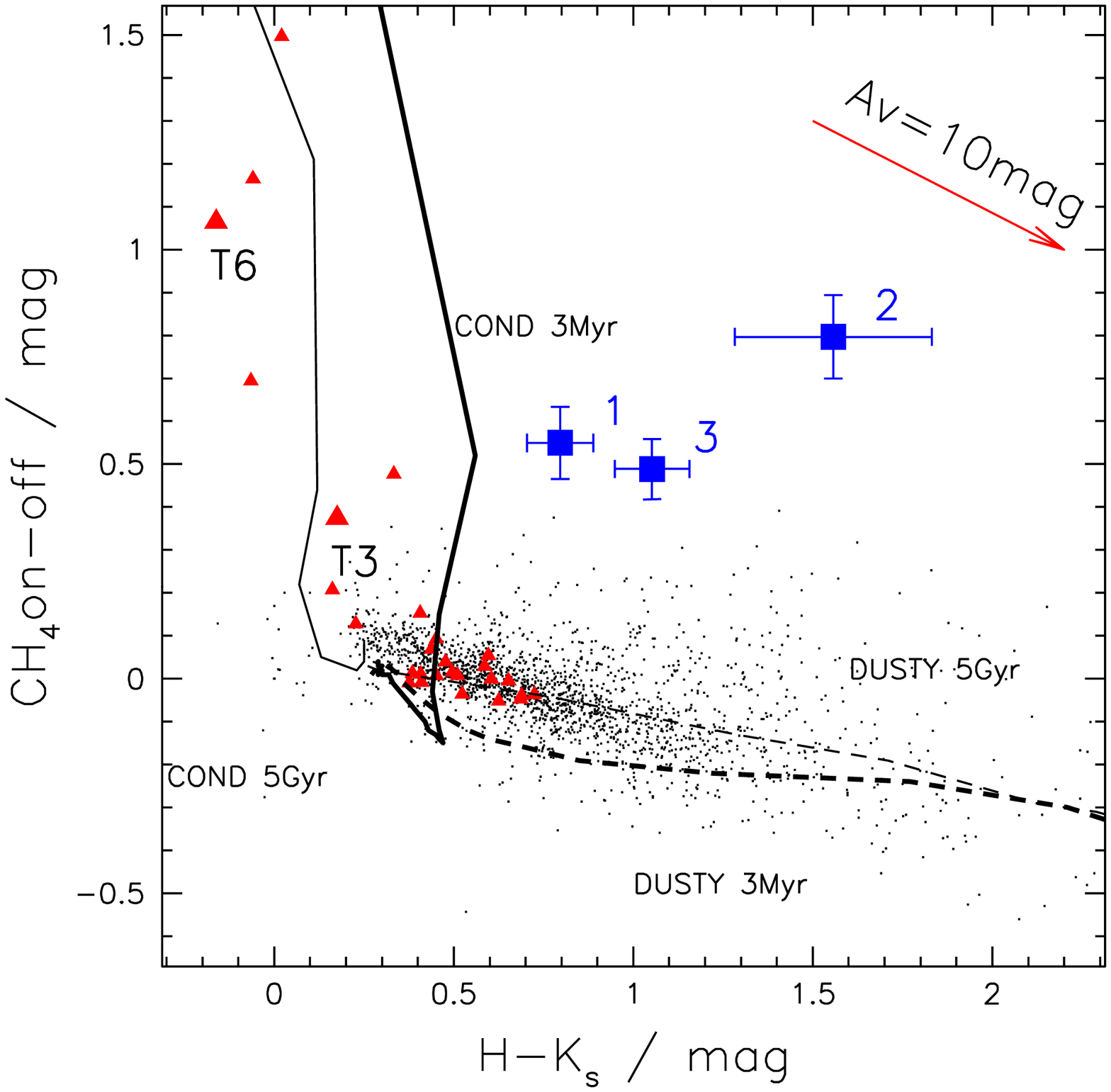} \\
& \\
\end{tabular}
\caption[$J-H$, $J-K_s$ and $H-K_s$ vs
  $CH_4$on$-CH_4$off]{$CH_4$on$-CH_4$off vs. $J-H$, $J-K_S$ and
  $H-K_S$. T-dwarf candidates are plotted as (blue)
  squares. Photometric error bars are from SExtractor. Field T-dwarfs
  (red triangles) are shown for comparison. Lines show the 5~Gyr COND
  and DUSTY, and 3 Myr COND and DUSTY models as in
  Fig.~\ref{figteff}. Candidates were dereddened towards the 3~Myr
  COND model using the extinction vector. }
\label{figJHOO}
\end{figure*}

Estimation of the candidates' extinction is required in order to be
able to calculate their absolute magnitude and estimate their spectral
type. Extinction has been estimated using colour/colour diagrams of
$CH_4$on$-CH_4$off versus $J-H$, $J-K_s$ and $H-K_s$, as plotted in Figure
\ref{figJHOO}. The extinction was computed for each candidate using
the extinction vector and regressing the objects back towards the 3
Myr COND model. The final extinction value is the average of the 3
results obtained from each colour/colour diagram and is summarised in
Table \ref{table:top3}, along with $CH_4$on$-CH_4$off colour, detection level and spectral type for the three candidates.  The estimated extinction is an upper limit if
the candidates belong to the cluster as their true dereddened colours
are probably intermediate between the 3~Myr COND and DUSTY models. If
they are field dwarfs however, the given value is a lower limit as the
objects should be dereddened towards the 5~Gyr field dwarf sequence
that is bluer than the 3~Myr COND model.

\begin{table*}[t]
\centering
\caption{Summary of values for the three candidates, where the cluster centre is taken to be
  at 03$^{\mathrm{h}}$44$^{\mathrm{m}}$34$^{\mathrm{s}}$
  +32$^{\circ}$09$^{\prime}$48${\farcs}$0 (J2000). }
\label{table:top3}

\begin{tabular}{l l l l l l l} 
\hline\hline 
IAU name & $CH_4$(on-off) & Det. & A${_V}$/mag & Est. sp. & Distance from & Object coordinates\\
&(mag)&level&&type& cluster centre & \\
\hline
\object{CFHT\_J0344+3202\_(IC348\_CH4\_1)} & 0.54&4.3 $\sigma$& 5.0 $\pm$ 1.2 & T5$^{+0.5}_{-1}$ & 7\arcmin35\arcsec  &  03$^{\mathrm{h}}$44$^{\mathrm{m}}$49$^{\mathrm{s}}$.24 +32$^{\circ}$02$^{\prime}$48${\farcs}$4             \\
\object{CFHT\_J0344+3206\_(IC348\_CH4\_2)} & 0.78&6.2 $\sigma$& 12.4 $\pm$ 3.9 & T6$^{+1}_{-0.5}$ & 4\arcmin20\arcsec &  03$^{\mathrm{h}}$44$^{\mathrm{m}}$49$^{\mathrm{s}}$.52 +32$^{\circ}$06$^{\prime}$35${\farcs}$4              \\
\object{CFHT\_J0344+3156\_(IC348\_CH4\_3)} &0.49 &4.3 $\sigma$&  9.0 $\pm$ 1.3 & T5 $\pm$ -0.5  & 13\arcmin56\arcsec &  03$^{\mathrm{h}}$44$^{\mathrm{m}}$57$^{\mathrm{s}}$.95 +31$^{\circ}$56$^{\prime}$43${\farcs}$3             \\
\hline 
\end{tabular}            

\end{table*}

We then used these results to compute the dereddened
$CH_4$on$-CH_4$off colour for each candidate. We found these values to
be 0.69 $\pm$ 0.16, 1.15 $\pm$ 0.24 and 0.76 $\pm$ 0.15~mag for
\object{IC348\_CH4\_1}, \object{IC348\_CH4\_2} and \object{IC348\_CH4\_3}, respectively. From
Figure~\ref{figspt}, these dereddened colours correspond to a spectral
type of T5$^{+0.5}_{-1}$, T6$^{+1}_{-0.5}$, and T5$^{+0.5}_{-0.5}$ for
\object{IC348\_CH4\_1}, \object{IC348\_CH4\_2} and \object{IC348\_CH4\_3}, respectively. As noted
above, this may be a lower estimate as the models suggest that young
T-dwarfs have a lower effective temperature than field T-dwarfs for
the same $CH_4$on$-CH_4$off colour (see Fig.\ref{figteff}). Moreover,
if the candidates were 5~Gyr field objects, they would have a larger
extinction and therefore an even larger intrinsic $CH_4$on$-CH_4$off
colour, which would also yield a later spectral type.


\section{Discussion}

We discuss here the likelihood that the methane candidates reported
above are {\it bona fide} young, very low-mass members of the \object{IC~348}
star-forming region instead of being more evolved field T-dwarfs
located on the line of sight to the young cluster. We also compare the
number of T-dwarf candidates we identify in our survey to the number of
expected planetary mass objects in \object{IC~348}, by extrapolating recent
estimates of the substellar IMF to the planetary mass domain.

\subsection{Membership}

\begin{figure}[t]
\setlength{\unitlength}{1cm}
\centering
\graphicspath{{images/}}
\includegraphics[width=1.0\linewidth]{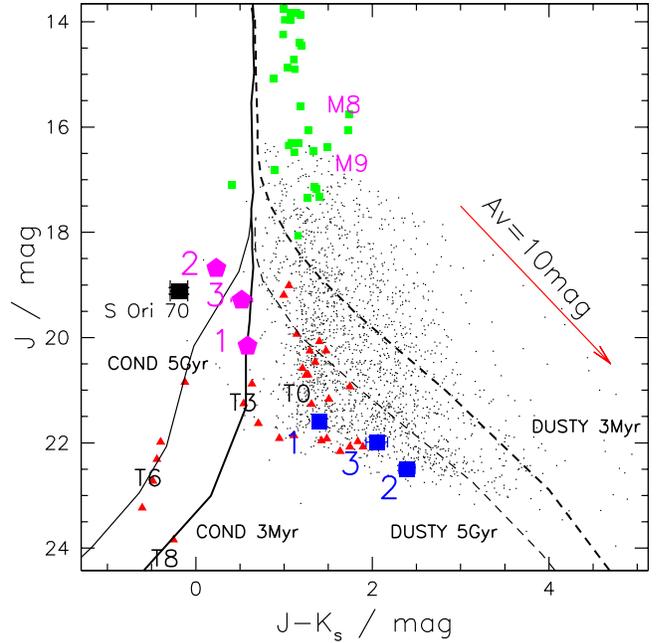}
\caption[$J$ vs $J-K_s$ ]{$J$ vs $J-K_s$ CMD. The (blue) squares are the
  three T-dwarf candidates as observed, whilst the (magenta) pentagons
  represent their dereddened location in the CMD. Field T-dwarfs (red
  triangles) as well as \object{S Ori 70} (black square, labelled) shifted to
  the distance of \object{IC~348} are shown for comparison. Dereddened M6-M9 IC~348 dwarfs from \cite{2003ApJ...593.1093L} are plotted for comparison (small green squares). DUSTY and COND
  models are shown as in Fig.~\ref{figteff}.}
\label{figJJK}
\end{figure}

Further colour/colour and colour/magnitude diagrams (CMD) were plotted
to constrain the candidates' status. Figure~\ref{figJJK} shows the
$J/J-K_s$ CMD. Also plotted are the synthetic magnitudes and colours of
the field T-dwarfs when shifted to the cluster distance. The 5 Gyr
DUSTY and COND models are drawn to highlight the agreement between the
models and the field dwarf sequence. The field T-dwarf sequence begins
in between the 5 Gyr DUSTY and COND models, before sweeping towards
the 5 Gyr COND model at $\sim$1500~K that sharply increases in
faintness with the late-T set.

In this diagram the three candidates appear to be confused with the
late-L, early-T field dwarfs. However, once they are dereddened using
the extinction values given in Table~\ref{table:top3}, they become
bluer and brighter. All three candidates appear to follow the 3~Myr
COND model quite closely. The fact that the dereddened candidates
appear brighter than field T-dwarfs shifted to the cluster distance
suggest that they are younger indeed. A young T-dwarf has a larger
radius than a field T-dwarf for a given spectral type as it is still
contracting. According to the COND models, the difference amounts to
about a factor of 2 in radius between 3~Myr and 5~Gyr (for
T$_{eff}$=1000-1500~K), which results in a 1.5 magnitude increase,
consistent with the observed location of the candidates in
Figure~\ref{figJJK}.

There is a possibility that the candidates are in fact field T-dwarfs
located at a closer distance (between $\sim$100 and 200 pc) along the line of sight to \object{IC~348}.  The probability of one of the candidates being a field T-dwarf instead
of an \object{IC~348} cluster member can be estimated
independently from extinction.  According to Figure~\ref{figspt}, the spectral type range T3-T5.5 corresponds to the
measured $CH_4$on$-CH_4$off colours of our candidates at an age of
5~Gyr.  Thusly, using the number density of T3-T5.5
field dwarfs in the solar neighbourhood, i.e. $\sim$1 per 714 pc$^3$
according to \cite{2008ApJ...676.1281M} and the footprint of
the $CH_4$ image of 0.11 sq.deg.
gives an estimate of 0.11 $\pm$ 0.06 T3-T5.5 foreground field dwarfs in the
direction of \object{IC~348}.  As we are concerned here with the probability of one of our
candidates being a foreground field dwarf,
this is a fairly robust method for determining the population density
for field dwarfs in the direction of \object{IC~348}.  However, this estimate is put into context when taking into account the large extinction values of the three candidates estimated from
Figure~\ref{figJHOO}.  Even the least extincted candidate, \object{IC348\_CH4\_1}, has four
magnitudes more extinction than expected for a foreground field dwarf
({\textless}1 mag), so all three objects must be near to or behind \object{IC~348}.  However, the candidates cannot be background field
T-dwarfs, seen through the \object{IC~348} cloud, as their luminosity would
then be much too high for their estimated spectral type.  Finally, as indicated in Table~\ref{table:top3}, all three candidates are
located within the cluster's boundary (4{\arcmin} core radius, 
10-15{\arcmin} halo; Herbig 1998, Herbst 2008) and are thus spatially
consistent with being \object{IC~348} members.

\subsection{Contaminants}

\begin{figure}[t]
\setlength{\unitlength}{1cm}
\centering
\graphicspath{{images/}}
\includegraphics[width=1.0\linewidth]{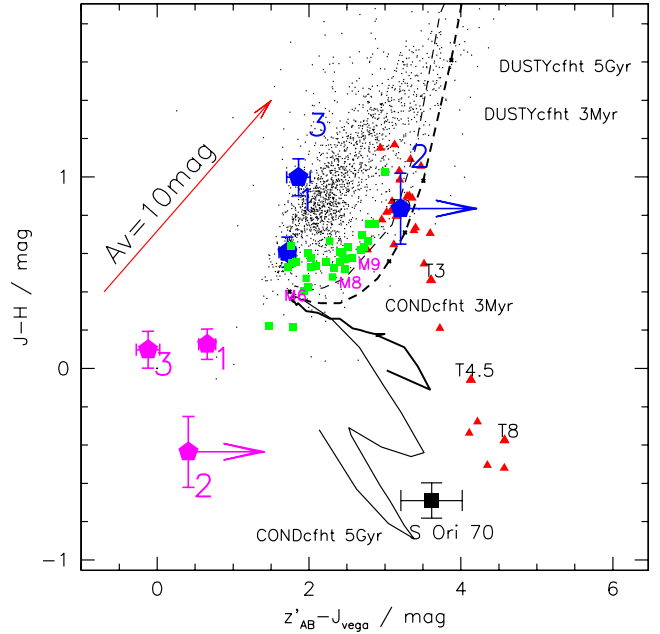}
\caption[$z'-J$ vs $J-H$]{$z'-J$ vs $J-H$ diagram.  DUSTY and COND
  models are shown as in Fig.~\ref{figteff}.  Symbols shown as per Fig.~\ref{figJJK}.}
\label{figzJJH}
\end{figure}

A further, useful diagram in defining these candidate objects is the $z'-J$/$J-H$ colour/colour diagram shown in Figure~\ref{figzJJH}, (note that the $z^{\prime}$ is in the AB system). Here it can be seen that the two detected dereddened candidates are much bluer in $z'-J$ than both the field dwarf sequence and the COND 3Myr and 5Gyr models.  There is mounting empirical evidence for young T-dwarfs to have bluer $z'-J$ colours than field T-dwarfs because of the effects reduced gravity has on the opacities (P. Delorme, priv. comm.).  This effect  stems from the strong potassium KI
(7687\&7701~$\mu$m) doublets, whose wings fall within the $z'$ band.  The lower the gravity the lesser the line broadening of these elements so less flux is lost by absorption in the $z^{\prime}$ band, resulting in the observed bluer $z'-J$ colours (F. Allard, priv. comm). However, these effects are unlikely to explain the extreme blueness of two of the three objects and suggests that candidates \object{IC348\_CH4\_1} and \object{IC348\_CH4\_3} are unlikely to be T-dwarfs.  The final remaining candidate, \object{IC348\_CH4\_2}, remains a good T-dwarf candidate because of its non-detection in $z'$. 

If candidates \object{IC348\_CH4\_1} and \object{IC348\_CH4\_3} are not young T-dwarfs then what are they? We checked whether extragalactic objects could contaminate this region of the diagrams. To this end, galaxies\footnote{iras.ipac.caltech.edu} from SWIRE, 2MASS and SDSS were found to have very different colours ($z'-J\sim$ 6-8mag) to our candidates.  Similarly, quasars \citep{2008MNRAS.383.1513L} also have J-band magnitudes of $\sim$~24-25 mag, making them much fainter than our objects.  Late-type (emission line) galaxies appear tightly in the region $J-K \sim$ 0.95 mag and $J-H \sim$ 0.88 mag and so again cannot be confused with our candidates \citep{2006MNRAS.372..199C}.  The status of the two rejected candidates continues to be unclear.

\subsection{Comparison of IC348\_CH4\_2 with S Ori 70}

\cite{2002ApJ...578..536Z} found \object{S Ori 70} to be a faint mid-T
type object towards the direction of the young $\sigma$ Orionis
cluster, with an estimated mass in the interval 2-7
M$_{Jup}$. However, there has been some contention with regard to the
membership of \object{S Ori 70} to the $\sigma$ Orionis
cluster. Further work by \cite{2008ApJ...672L..49S} indicated that
this was a true cluster member by looking at the IRAC data, and
suggested that there is a possibility of a disk structure which gives
rise to the MIR excess.

The $J/J-K_s$ CMD in Figure~\ref{figJJK} has the
IC~348 dereddened candidates plotted alongside \object{S Ori 70} for
comparison. The $J, H$ and $K_s$ data for \object{S Ori 70} was taken
from \cite{2008A&A...477..895Z} and adjusted to the CFHT photometric
system using the 3~Myr COND CFHT and 2MASS models. The differences for
$\lambda_{2MASS}-\lambda_{CFHT}$ due to colour effects are 0.6 mag,
-0.1 mag, and -0.05 mag, for $J$, $H$ and $K_s$ respectively. In this
diagram, candidate \object{IC348\_CH4\_2}, for which we estimated a spectral type of
T6$^{+1}_{-0.5}$, falls at approximately the same location as
\object{S Ori 70}, shifted to \object{IC~348}'s distance.

\begin{figure}[t]
\setlength{\unitlength}{1cm}
\centering
\graphicspath{{images/}}
\includegraphics[width=1.0\linewidth]{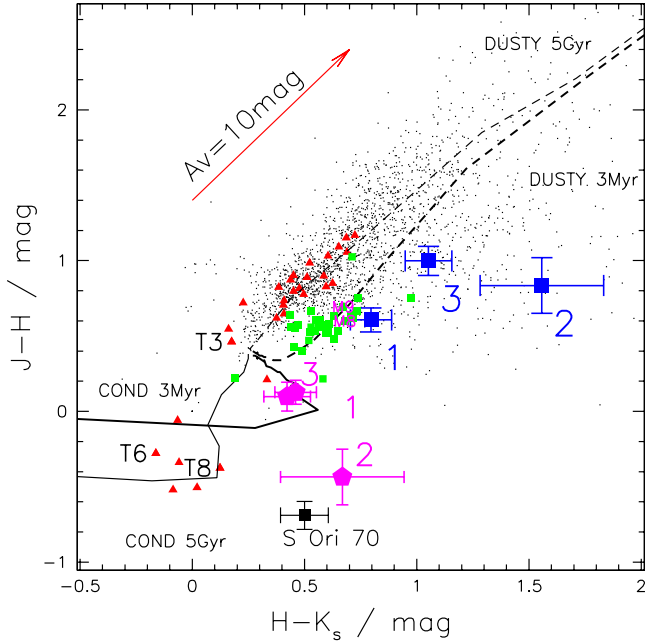}
\caption[JHvsHK]{$J-H$ vs $H-K_s$. Symbols as per Fig.~\ref{figJJK}. }
\label{figJHHK}
\end{figure}

Figure~\ref{figJHHK} shows the $J-H/H-K_s$ colour/colour diagram. The dereddened $J-H$ colours of remaining \object{IC348\_CH4\_2} T-dwarf candidate is -0.44 which corresponds to a spectral type of around T6-T8 \citep{2007ApJ...654..570L} and so is consistent with our previous estimate.  Again in this plot, the
location of \object{IC348\_CH4\_2} is similar to that of \object{S Ori 70}.  A remarkable
property of these two objects is that they have $H-K_s$ colours too
red for their estimated spectral types, yet also have blue $J-H$
colours consistent with them being T6 dwarfs.  According to
\cite{2008A&A...477..895Z}, this may be a signature of youth.

\subsubsection{Spitzer data}
We searched the {\it Spitzer} public archive for IRAC data. IC~348 was
the prime target of programmes 6 (c2d, P.I. Fazio) and 36 (P.I. Fazio).
The c2d data are made up by shorter exposures, resulting in our use of the deeper
(16$\times$100s) images of programme 36. We retrieved the data and
processed them using standard procedures with the recommended MOPEX
software. All three candidates are detected in the $[3.6]$ and $[4.5]$ bands, except
IC348\_CH4\_1 which falls out of the $[4.5]$ band field of view. PSF
photometry was performed using Starfinder \citep{2000A&AS..147..335D} and the
fluxes were translated into magnitudes using the zeropoint fluxes
provided by the Spitzer Science Center\footnote{http://ssc.spitzer.caltech.edu/}. Measurement uncertainties were tentatively
estimated from the Poisson noise weighted by the coverage maps of the
mosaics. The final photometry is given in Table \ref{table:IRACdata}, and the final
errors include both measurement and zeropoint flux
uncertainties.

\begin{table*}[t]
\caption{ Spitzer PSF photometry and photometric errors of the three T-dwarf candidates. IC348\_CH4\_1 appeared outside the FOV in the $[4.5]\mu$m image.}
\label{table:IRACdata}  
\centering 
\begin{tabular}{l l l l l l l l l}
\hline\hline
 Object  &       F${[3.6]}$/mJy & F${[3.6]}$\_error & F${[4.5]}$/mJy & F${[4.5]}$\_error & [3.6]/mag & [3.6]\_error & [4.5]/mag & [4.5]\_error \\
\hline
 \object{IC348\_CH4\_1} & 0.0035 & 0.0001 & - & - & 19.75 & 0.03 & - & - \\ 
 \object{IC348\_CH4\_2} & 0.0093 & 0.0003 & 0.0136 & 0.0004 & 18.70 & 0.03 & 17.80 & 0.03\\
 \object{IC348\_CH4\_3} & 0.0245 & 0.0003 & 0.0365 & 0.0007 & 17.65 & 0.02 & 16.73 & 0.02\\ 
\hline                                                                                           \end{tabular}
\end{table*}

\begin{figure}[htbp]
\setlength{\unitlength}{1cm}
\centering
\graphicspath{{images/}}
\includegraphics[width=1.0\linewidth]{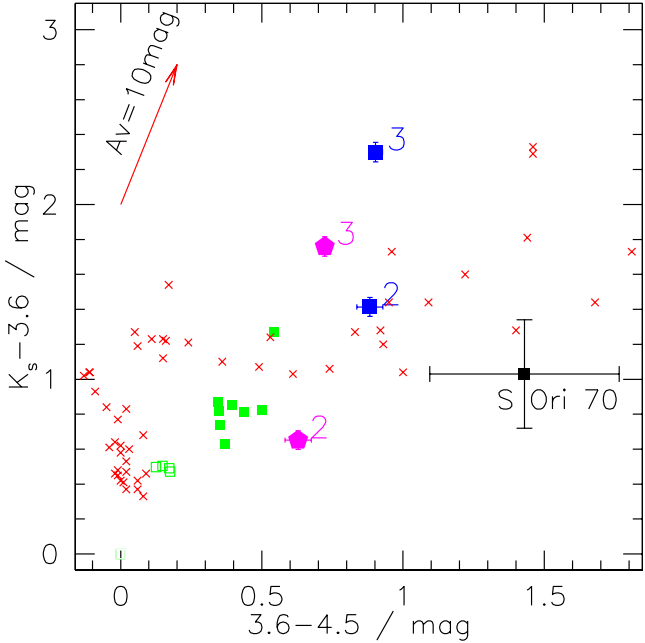}\\
\caption[$K_{s}-3.6$ vs $3.6-4.5$]{$K_{s}-[3.6]$ vs $[3.6]-[4.5]$.  The field dwarf sequence from \cite{2006ApJ...651..502P} has been plotted (red crosses) for comparison. \object{IC~348} M6-M9 dwarfs with disks (filled green squares) and those without (hollow green squares) are plotted \citep{2005ApJ...631L..69L}.  \object{IC348\_CH4\_2} and \object{S Ori 70} appear significantly bluer in the $K_{s}-[3.6]$ colour than for field dwarfs.  The extinction vector has also been plotted using values from the Spanish Virtual Observatory\footnotemark[3].} 
\label{figIRAC}
\end{figure}

In Figure~\ref{figIRAC}, the $K_{s}-[3.6]$ vs $[3.6]-[4.5]$ colour/colour diagram is plotted with the two candidates, S Ori 70 and the \object{IC~348} M6-M9 dwarf sequence from \cite{2005ApJ...631L..69L}.  The M, L and T IRAC field dwarf sequence from \cite{2006ApJ...651..502P} have also been plotted for comparison. The $[3.6]-[4.5]$ colour could be a good indicator of effective temperature, or spectral type, giving \object{IC348\_CH4\_2} a spectral type of $\sim$T5, in line with our other estimates of T6.  The $K_{s}-[3.6]$ colour, however, relates to a spectral type of L1, which is far too early for such a low-mass and cool object.

The $K_{s}-[3.6]$ colour of \object{IC348\_CH4\_2} appears significantly bluer compared to the field sequence, which may be because of reduced gravity \citep{2007ApJ...655.1079L}.  For mid-type and later T-dwarfs the reduced pressure broadening of H$_2$, as a gravity effect, makes the $K_s$ band brighter, whilst the $[3.6]$ band faintens because of additional CH$_4$ absorption at $\sim$ 3 $\mu$m, resulting in a bluer $K_{s}-[3.6]$ colour.  Another consequence of lower gravity is the faintening effect the CO abundances have on the $[4.5]$ band for early T-types, thus balancing the $[3.6]-[4.5]$ colour. These colour effects of low gravity are in agreement with this object being a young IC 348 T-dwarf. 

Other effects such as strong sedimentation and little or no vertical mixing in the atmosphere could also contribute to the observed bluer $K_{s}-[3.6]$ colours \citep{2007ApJ...655.1079L}.

The possible existence of a hot inner dusty disc around S Ori 70, in keeping with other young members of S Orionis, can give rise to its redder $[3.6]-[4.5]$ colour.

\subsection{The lower end of the IMF}

\cite{2003ApJ...593.1093L} derived a nearly complete IMF for \object{IC~348}
down to 0.03~M$_{\sun}$ for A$_V\leq$~4~mag.  They find a ratio of
brown dwarfs (BDs, 0.02-0.08~M$_\odot$) to stars of about 12\%.
\cite{2003AJ....125.2029M} similarily derived an IMF for \object{IC~348} down
to 0.04~M$_\odot$, with a mode between 0.08 and 0.2~M$_\odot$, finding
a ratio of BDs to stars of about 14\%.  From a deep J-band survey of
the cluster, \cite{2003A&A...409..147P} derived a BD to star ratio of
10\%.

According to the COND and DUSTY models, the mass range of 3~Myr old
T-dwarfs is between 0.001~M$_{\sun}$ and 0.005~M$_{\sun}$,
corresponding to masses from $\sim$1~M$_{Jup}$ to $\sim$5~M$_{Jup}$.  However, the models are somewhat uncertain at very low
masses and young ages. Currently the models can have up to a 50\%
discrepancy in the masses for objects younger than 5~Myr (Chabrier,
priv. comm).  Still, the T-dwarf candidate \object{IC348\_CH4\_2} is likely to be less massive
than 10 M$_{Jup}$ if belonging to \object{IC~348}.

In order to obtain an estimate of the number of objects within this
mass range that are expected to be in \object{IC~348}, we extrapolated current
estimates of the IMF to the planetary mass regime.  A lognormal
estimate of the field IMF for unresolved systems was provided by
\cite{2003PASP..115..763C}, where the mode of the distribution, $m_0$,
is 0.22 M$_{\sun}$ and its width, $\sigma$, is 0.57.  This lognormal
IMF would predict $\sim$1\% objects in the mass range below
10~M$_{Jup}$. Similarily, \cite{2007A&A...471..499M} suggested a
universal lognormal IMF for systems in young open clusters, with
$m_0$=0.30 $\pm$ 0.05 M$_{\sun}$ and $\sigma$=0.55 $\pm$ 0.03, which
would predict $\sim$0.4\% of objects in the 1-10~M$_{Jup}$ mass range. \cite{2007AJ....134..411M} conducted a Spitzer census of \object{IC~348} and
stated that the population of \object{IC~348} is in excess of 400 members when
taking into account unseen diskless members.  Based on the above IMF and population estimates, we would thus expect about 1.6-4 objects in the mass range 1-10~M$_{Jup}$ in the cluster.  The discovery of one T-dwarf candidate close to the completeness limit is thus consistent with the extrapolation of current lognormal IMF estimates down to the planetary
mass domain.

Another estimate can be found using the power law derived by \cite{2007A&A...470..903C} and \cite{2009arXiv0907.2185L} for \object{$\sigma$ Orionis} and \cite{2007MNRAS.374..372L} for \object{Upper Sco}, $\frac{dN}{dM} \propto M^{-\alpha}$, where $\alpha$ = 0.6 for the mass range $0.3-0.01$~M$_{\sun}$.  Using the IMF from the A$_V\leq$~4~mag selected census of \object{IC~348} by \cite{2003ApJ...593.1093L} and extrapolating the power law for the last two data points above their completeness limit towards the lowest mass domain gives a different estimate.  This corresponds to a much larger number of predicted objects than for the lognormal estimate, 25 $\pm$ 16 objects in the 1-10~M$_{Jup}$ mass range and 15 $\pm$ 9 objects in the 1-5~M$_{Jup}$ mass range.  Even though our survey is complete down to A$_V\sim$~12~mag for T3 - T5.5 dwarfs, we did not detect any such object with an A$_V\leq$~4~mag.  This suggests that the current power law approximation over-estimates the number of low-mass objects in \object{IC~348}.

\section{Conclusions}

From a deep methane imaging survey of the star-forming region \object{IC~348}
we identified 3 T-dwarf candidates over the area of the cluster. After colour/colour and colour/magnitude diagram analysis two candidates have been rejected for being too bright at optical wavelengths.  The remaining candidate, has
an estimated spectral type of T6 and theoretical models suggest a mass of a few M$_{Jup}$ for this object at 3~Myr.  From its luminosity, colour, extinction and spatial
location, \object{IC348\_CH4\_2} is a probable \object{IC~348} T-dwarf member, and so
is among the lowest mass objects observed so far in a star-forming
region.  The frequency of
isolated planetary mass objects reported here for \object{IC~348} is consistent
with the extrapolation of current lognormal IMF estimates to the planetary mass
domain.

\begin{acknowledgements}
We thank F. Allard for many discussions during the writing of this paper, and also to the referee for providing insightful comments with regard to this work. Many thanks goes to P. Delorme for his useful comments regarding cool dwarfs. We thank also Manuel Perger and David Barrado y Navascu\'{e}s for discussions and Subaru data. We thank the QSO
team at CFHT for their efficient work at the telescope and the data
pre-reduction as well as the Terapix group at IAP for the image
reduction. This work is based in part on data products produced and
image reduction processes conducted at TERAPIX. This research has made
use of the NASA/ IPAC Infrared Science Archive, which is operated by
the Jet Propulsion Laboratory, California Institute of Technology,
under contract with the National Aeronautics and Space
Administration. This research has also made use of the SIMBAD
database, operated at CDS, Strasbourg, France.
\end{acknowledgements}

\bibliographystyle{aa}
\bibliography{12444.bbl}

\end{document}